\documentstyle[prb,aps,twocolumn]{revtex}

\input epsf

\def\prb{Phys.\ Rev.\ {\bf B}}
\def\prl{Phys.\ Rev.\ Lett.\/}

\renewcommand{\c}[2]{C_{\vec #1,#2}}
\newcommand{\cd}[2]{C^{\dagger}_{\vec #1,#2}}
\newcommand{\be}{\begin{equation}}
\newcommand{\ee}{\end{equation}}
\newcommand{\n}[2]{n_{\vec #1}^{#2}}
\newcommand{\s}{\sigma}
\newcommand{\ba}{\begin{eqnarray}}
\newcommand{\ea}{\end{eqnarray}}
\renewcommand{\sp}[2]{\vec{S_{#1}}\cdot\vec{S_{#2}}}
\newcommand{\qu}{{1\over 4}}
\newcommand{\p}[1]{S_{#1}}

\begin{document}

\twocolumn[\hsize\textwidth\columnwidth\hsize\csname@twocolumnfalse\endcsname

\title
{Moments of the ARPES spectral function of an undoped Mott
insulator.}

\author{Vadim Oganesyan}
\address
{Dept. of Physics,
University of California,
Los Angeles, CA  90095}
\date{\today}
\maketitle 

\begin{abstract}

\
We derive analytic expressions for the first three frequency moments of the single
particle spectral function for one hole in a Mott insulator in terms of
equilibrium spin correlation functions of the insulating state.  We show
that the ``remnant Fermi surface`` detected in ARPES experiments is, in
fact, a reflection of the strong antiferromagnetic correlations of the
system, {\it not} a reflection of the original band-structure Fermi
surface.  We suggest that ARPES data could be used to measure the magnetic
specific heat.
\
\end{abstract}
]
\

In this paper we analyze the single hole spectral function of a Mott 
insulator.  We take as our model of the 
insulator the large U Hubbard model at half-filling and find that spectral 
moments can be evaluated using standard perturbation 
theory in the kinetic energy.  Equilibrium magnetic correlations of the half-filled state fully
characterize the one hole spectral moments.

Even well understood strongly correlated electron systems 
have non-trivial single particle properties.  Loosely 
speaking, when the effects of 
interactions, kinematics, and/or magnetic field overwhelm the kinetic 
energy of the electrons, the resultant new states of matter bear little 
or no resemblance to the non-interacting electron gas.  And while sharp single particle 
spectral peaks are the defining property of weakly correlated systems (metals, 
semiconductors, even BCS superconductors) broad incoherent 
features, often alongside a {\it weak} quasi-particle-like (QP) peak, are more common in 
strongly correlated materials.  Specifically, angular resolved photoemission
 spectroscopy 
(ARPES) data taken from Ca$_{2}$CuO$_{2}$Cl$_{2}$ (a prototype Mott insulator) does 
show\cite{zx} a 
peak in a limited region of phase space but is clearly dominated by broad features.  Numerical solutions on small systems 
combined with judiciously summed perturbation schemes successfully reproduce some aspects of photoemission results\cite{ed,qmc,scba}.  However the
underlying physics determining the spectral function 
is often obscured; even more often the approximation scheme is not 
universally accepted as justifiable.
In contrast, by asking different (though related) questions we have not only drawn a clear
connection between one hole properties with those of the undoped antiferromagnet (AF), but have done so in a
controlled manner.  The basis of our 
calculation will be reviewed in Section 2.  In Section 3 we will present and discuss our result for
the zeroth moment, $\n{k}{}$.  In particular, we will show that ``a remnant Fermi surface,'' defined as
either the locus of points in $\vec k$ space at which $\n{k}{}=1/2$ or as the locus of points of maximum
gradient of $\n{k}{}$, lies close to the ``diamond Fermi surface'' ($\cos(k_x)+\cos(k_y)=0$),
{\it regardless} of whether or not this is the location of the Fermi surface in the $U\rightarrow 0$ limit. 
Results concerning higher moments are presented in Section 4.   Some
of the technical details can be found in the Appendix.

\section{The Calculation}
\subsection{Half-filled Hubbard model at large U}

The large U Hubbard model at half filling is the simplest
model that 
produces the large charge gap and the antiferromagnetic tendencies observed 
experimentally in CuO planes.
The Hamiltonian,
\be
\hat{H}=-\sum_{\s,i,j}t_{ij}\cd{i}{\s}
        \c{j}{\s}+U\sum_{\vec{j}}\n{j}{\s}\n{j}{-\s}
\ee
contains a strong onsite repulsion term $U\sum_{\vec{j}}\n{j}{\s}\n{j}{-\s}$ 
that freezes (together with the lattice potential) charge motion at half-filling provided that
the kinetic energy term is small enough to be treated as a 
perturbation, $|t_{ij}|\ll U$.

When $t_{ij}/U=0$ the ground state consists of singly occupied sites, 
has energy zero, is $2^{N}$-fold degenerate where N is the size of the system,  and
is completely characterized by the spin configuration, $\{ S_{i} \}$.

The perturbative effects of the kinetic energy modify all of the above statements except the last
one: each of the perturbed states, though it contains an admixture of doubly occupied sites, can still be
labeled by the spin configuration of the unperturbed state from which it has evolved.  Thus the expectation
value of {\it any} operator in the ground state manifold can always be expressed in terms of spin
variables.
Formally this is accomplished by computing perturbatively in powers of $t_{ij}/U$ the unitary transformation 
$\exp[i\hat{X}]$ that expresses the evolution of the low energy unpeturbed states $|\{ S_{i} \} >$ as a
function of increasing kinetic energy\cite{mdgy}:
\ba
\widetilde{|\{ S_{i} \} >} &=& e^{-i\hat{X}}| \{ S_{i} \} >\\
< \hat{\mathcal{O}} > &=& < \{ S_{i} \} |e^{i\hat{X}}\hat{\mathcal{O}}e^{-i\hat{X}}| \{ S_{i} \} >,
\ea
where $\mathcal{O}$ is any observable.

In the familiar fashion, this transformation maps the low energy physics of the Hubbard Hamiltonian 
into an effective Heisenberg antiferromagnet whose  leading term is:
\ba
\hat{H}_{eff} &=& e^{i\hat{X}}\hat{H}e^{-i\hat{X}} \nonumber \\
            &=& -\frac{1}{2}\sum_{i,j} \frac{4t_{ij}^{2}}{U}(\qu-\sp{i}{j}) +{\cal O}(t_{ij}^4/U^3).
\ea

\subsection{The spectral function}
The emission spectral function is 
\ba
A(\vec{k},w)&=&\frac{1}{2}\sum_{\s,m,n}
e^{-\beta E_n}|<m|\cd{k}{\s}|n>|^2\delta(\omega+E_{m}-E_{n}) \nonumber \\
&=&\frac{1}{2}\sum_{\s}\int dt e^{iwt}<\cd{k}{\s}(t)\c{k}{\s}(0)>.
\ea

Frequency moments of $A(\vec{k},w)$ correspond to ground state (or thermodynamic)
averages of the following
operators: 
\begin{eqnarray}
\n{k}{}      &\equiv& \int\frac{dw}{2\pi}A(k,w) \ =\ \frac{1}{2}\sum_{\s} <\cd{k}{\s}\c{k}{\s}>\\
A_1(\vec{k})&\equiv& \int\frac{dw}{2\pi} \omega \ A(\vec{k},w) \nonumber \\
&=&-\frac{1}{2}\sum_{\s}<\cd{k}{\s}[\hat{H},\c{k}{\s}]>\\        
A_2(\vec{k}) &\equiv& \int\frac{dw}{2\pi}\omega^{2}A(\vec{k},w) \nonumber \\
				&=&\frac{1}{2}\sum_{\s}<\cd{k}{\s}[\hat{H},[\hat{H},\c{k}{\s}]]>
\end{eqnarray}

A systematic evaluation of these averages, using the perturbative expression for $\hat X$ in powers of
$t_{ij}/U$, can be found in the Appendix.

\section{The spectral weight, $\n{k}{}$}
We begin with a discussion of the spectral weight because it has an immediate physical
interpretation as the occupation probability.  As shown in the Appendix,
\be
\n{k}{}=\frac{1}{2}[1-\frac{4\tilde{\epsilon_{k}}}{U}+O(t^3/U^{3})]
\label{eq:n}
\ee
where
\be
\tilde{\epsilon_{k}}=-\sum_{j} \p{0i}  t_{0j} e^{i\vec{k} \cdot \vec{R}_{j} }
\ee
and 
\be 
\p{ij}=<\frac{1}{4}-\sp{i}{j}>
\ee
is the equilibrium
spin correlation between spins i and j.  $\tilde {\epsilon_{k}}$ is a sort of
renormalized band energy 
in which each hopping matrix element $t_{ij}$ is renormalized by a factor of 
$\p{ij}$.  However, this renormalized energy does not correspond in any simple way to the energy of any
elementary excitation of the system. Note that the non-interacting free electron band is given by $\epsilon_{k}=-\sum_{j}  t_{0j} e^{i\vec{k} \cdot \vec{R}_{j} }$.

This expression for $n_{\vec{k}}$ can be derived in a different, simpler manner, which is readily generalizible to more
complicated situations, such as the three band Cu-O or Emery model\cite{vje}.  From the Hellman-Feynman theorem, it
follows that
\be
\sum_{\sigma} <[c^{\dagger}_{i\sigma}c_{j\sigma}+ {\rm H.C.}]>=-\partial E/\partial t_{ij}
= [\partial J_{ij}/\partial t_{ij}]\p{ij}
\ee
where $E$ is the internal energy, which can be computed using the effective Hamiltonian $\hat H_{eff}$. 
From the expression for $J_{ij}$ in terms of $t_{ij}$, the result in Eq.\ref{eq:n}
follows immediately.

\subsection{The remnant fermi surface}
At low temperatures, the short-range spin correlation functions are essentially temperature independent, and
equal to their value in the ground state.  To be concrete, let us consider the
Hubbard model with nearest, second, and third neighbor hopping, $t$, $t^{\prime}$ and
$t^{\prime\prime}$, respectively;  this sort of model was used in the numerical studies to fit the
dispersions seen in ARPES. The zero temperature spin correlations of the corresponding spin 1/2 Heisenberg
model have been computed fairly accurately in numerical studies\cite{4by4}.  For the Heisenberg model with
only nearest-neighbor exchange coupling, the spin correlation functions are
$\p{01}\approx 7/12$ and
$\p{03}\approx\p{02}\approx\frac{1}{20}$ between nearest, next-nearest, and third nearest neighbor
sites, respectively.  These correlations are, moreover, found to be relatively insenstitive to the inclusion
of a modest amount of further neighbor exchange couplings, which anyway are expected to be quite small since
$J^{\prime}/J = [t^{\prime}/t]^2$. 

In computing  $\tilde{\epsilon}_{k}$ the antiferromagnetic correlations between neighboring spins
imply a factor of 1/2 renormalization of $t$, compared to a factor of 1/20 renormalization of $t^{\prime}$ 
and $t^{\prime\prime}$.  Since in most cases of physical interest, $|t'|,|t''| \ll 10|t|$, even when
$t^{\prime}$ and $t^{\prime\prime}$ are large enough to make signicant shifts in the original Fermi surface
defined by $\epsilon_{k}$, the occupation probability is well approximated (Figure 1) as
\be
\n{k}{}\approx\frac{1}{2}[1+\frac{7 t(cos(k_x)+cos(k_y))}{3 U}]
\ee
It is both $\frac{1}{2}$ and has the steepest slope along the Fermi surface of
the non-interacting electrons with only the nearest neighbor hopping. In fact, ARPES experiments that observe such momentum dependence of $\n{k}{}$  have been interpreted as an indication that there is a ``remnant Fermi surface''
in the undoped Mott insulator.  By contrast, our result suggests that the observed $\n{k}{}$ is reflective of the the spin physics of
the strongly correlated Neel state rather than a vestige of the original Fermi surface.



\begin{figure}
\begin{center}
\leavevmode
\epsfxsize=3in
\epsfbox{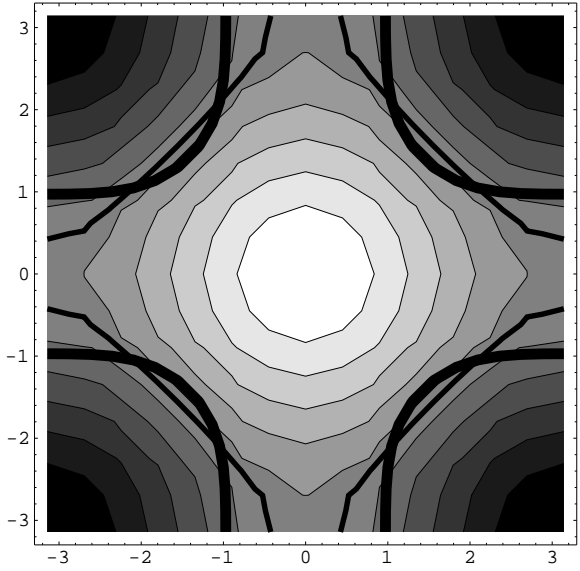}
\end{center}
\caption
{$\n{k}{}$ with $t'=-.3t, t''=.15t$ and spin correlations $\p{1}=7/12$ and $\p{2}=\p{3}=1/20$.Thick line is the 
non-interacting Fermi surface, while thin line is where $\n{k}{}=1/2$}
\label{fig:fig1}
\end{figure}


\subsection{Specific heat of the Neel transition}
The natural connection we find between the spectral weight and the Neel state can be exploited
further. Since, as we already pointed out, the Heisenberg Hamiltonian is dominated by the nearest neighbor
term, and assuming that $J_{n.n.}$ is only weakly (if at all) temperature dependent, the specific heat is
\be
C(T)=J\frac{\partial}{\partial T} <\frac{1}{4}-\sp{1}{0}>  = J\frac{\partial \p{1}}{ \partial T}.
\ee
Since
\be
\n{k}{}\approx\frac{1}{2}[1+\frac{8t(cos(k_x)+cos(k_y))}{U}\p{1}],
\ee
by measuring the temperature dependence of $\p{1}$ (as extracted from $\n{k}{}$) and differentiating it
one gets the magnetic contribution to the specific heat.
The specific heat extracted in this manner contains {\it only} the contribution from the
spin fluctuations.

\section{Higher Moments}
Higher spectral moments correspond to the derivatives $<\cd{k}{}(t)\c{k}{}(0)>$  at $t=0$ and thus provide further insight into the problem of a hole in a Mott insulator.  Our results for first and second moments, $A_1(\vec{k})$ and $A_2(\vec{k})$, are presented in a table below.  The moments are written as sums over momentum space Fourier harmonics, each corresponding to a summation over sites a given near neighbor distance away on the square lattice.
\ba
A_n(\vec{k})&=&\sum_{j}A_{n}^{\ j}\gamma_{j}(\vec{k}),\nonumber \\
\gamma_{j}(\vec{k})&=&\sum_{\vec{R}=j^{th} n.n.}e^{-i\vec{k}\cdot\vec{R_{j}}}
\ea

\begin{tabular}{|c|c|c|}\hline
i& $A_{1}^{\ j}$				& $A_{2}^{\ j}$ \\ 
\hline \hline
1& $t( \frac{1}{2} - \p{1})$					& $-(2tt'+tt'')(\p{1}-1)$  \\
 &$--\frac{2t't''}{U}(\p{2}+2\p{1})$				&               \\
 &$-\frac{tt''}{U}(\p{3}+2\p{1})$				&               \\
\hline
2& $-t'( \frac{1}{2} - \p{2})$					&  $-(2t't''+t^{2})(\p{2}-1)$  \\
 &$-\frac{2t't''}{U}(\p{3}+2\p{2})$				&               \\
 &$-\frac{t^{2}}{U}(\p{2}+2\p{1})$				&               \\
\hline
3& $-t''( \frac{1}{2} - \p{1})$					& $-((t')^{2}+\frac{t^{2}}{2})(\p{3}-1)$ \\ 
 &$-\frac{(t')^{2}}{U}(\p{3}+2\p{2})$				&               \\
 &$-\frac{t^{2}}{2U}(\p{3}+2\p{1})$				&               \\
\hline  
4&$-\frac{tt'}{U}(\p{4}+\p{2}+\p{1})$     			&$-(t't''+tt'')(\p{4}-1)$         \\
 &$-\frac{tt''}{U}(\p{4}+\p{3}+\p{1})$    			&               \\
\hline  
5&$-\frac{(t')^{2}}{U}(\p{5}+2\p{2})$				&$ -(\frac{(t')^{2}}{2}+(t'')^2)(\p{5}-1)$ \\
 &$-\frac{(t'')^2}{U}(\p{5}+2\p{3})$				&               \\
\hline  
6&$-\frac{tt''}{U}(\p{6}+\p{3}+\p{1})$    			&$ -tt''(\p{6}-1)$       \\
\hline
7&$-\frac{t't''}{U}(\p{7}+\p{3}+\p{2})$   			&$ -t't''(\p{7}-1)$      \\
\hline
10&$-\frac{(t'')^{2}}{2U}(\p{10}+2\p{3})$  &$ -\frac{(t'')^{2}}{2}(\p{10}-1)$       \\
\hline 
\end{tabular}

\

\

One notices that all computed moments are finite.  Although the existance of the moment expansion is a general requirement of any physical system, it is more rule than exception that approximations lead to divergencies past some finite order.  
Though we haven't constructed an explicit proof, there are indications that the expansion is well behaved for the Hubbard model.

In principle, the moment expansion can be used to study the QP dispersion ($E_{\vec{k}}$) directly: a coherent oscillation results in $E_{\vec{k}}^n$ contribution to $A_{n}(\vec{k})$.  The crudest (single mode) approximation of this sort identifies $\bar{E}_{\vec{k}}=A_1(\vec{k})/n_{\vec{k}}$.  We found it to be in a surprising agreement with previously obtained results for 
the momentum dependence of the QP energy in t-t'-t''-J as well as t-J models (one needs to assume that term proportional to t does not contribute to QP dispersion).
Since  $\bar{E}_{\vec{k}}$ rather seriously overestimates the overall bandwidth, it isn't clear if one is justified claiming to have 
obtained even an approximate QP energy yet.  Another likely use of our results (or rather their extension to higher moments and 
orders in perturbation theory) can be in  comparing with spectral functions obtained by other means (either numerics, self-consistent 
Born approximation or other).

In conclusion, we have outlined a well controlled method for analysing the spectral moments of a hole in a Mott insulator.  We find the occupation probability, $\n{k}{}$, is in agreement with the well established experimental result, which as our calculation suggests is strongly constrained by the presence of AF order. We further propose that the temperature dependence of $\n{k}{}$ can be used to study the specific heat of the Neel transition.  The implications of our results for higher moments are yet to be understood properly.

I would like to thank Steven Kivelson for innumerable illuminating conversations and suggesting this problem in the first place.  I would also like to greatfully acknowledge M.Z.Hasan, A. Lanzara, F.Ronning and Z.X.Shen for useful discussions and the hospitality of the Physics and Applied Physics Departments at Stanford University where this work was carried out.
This work was supported by NSF grant number DMR98-08685.
\appendix
\section{Evaluation of moments}
To compute the moments we first express them in terms of real space electronic correlations
\ba
n_{k}{}&=&\frac{1}{2}\sum_{\s}<\cd{k}{\s}\c{k}{\s}> \nonumber \\
&=& \frac{1}{2}\{1+\sum_{i,\s}<\cd{i}{\s}\c{0}{\s}>\gamma_{i}(k)\} \\
A_1(\vec{k})&=&-\frac{1}{2}\sum_{\s}<\cd{k}{\s}[\hat{H},\c{k}{\s}]>\nonumber \\
&=&\epsilon_{k}n_{k}+\frac{U}{2}\sum_{i,\s}\gamma_{i}(k)<\cd{i}{\s}\c{0}{\s}n_{0,-\s}>\\
A_2(\vec{k})&=&\frac{1}{2}\sum_{\s}<\cd{k}{\s}[\hat{H},[\hat{H},\c{k}{\s}]]> \nonumber \\
&=&\epsilon_{k}\n{k}{}\bar{E}_{\vec k}+\frac{U^2}{2}\sum_{i,\s}\gamma_{i}(k)<\cd{i}{\s}\c{0}{\s}n_{0,-\s}> \nonumber \\
&+&\frac{U}{2}\sum_{i,j,\s}\gamma_{i}(k)t_{j,0}<\cd{i}{\s}(-\c{j}{\s}n_{0,-\s} \nonumber \\
&+&\c{j}{-\s}\cd{0}{-\s}\c{0}{\s}+\cd{j}{-\s}\c{0}{-\s}\c{0}{\s}>
\ea
As before $\gamma_{i}(\vec{k})=\sum_{R_{i}}e^{-i\vec{k}\cdot\vec{R_{i}}}$,where the sum is over i'th nearest neighbors.

Next, these electronic correlations are evaluated perturbatively ($\p{ij}=<\frac{1}{4}-\vec{S}_i\cdot\vec{S}_j>$):
\ba
\sum_{\s}<\cd{i}{\s}\c{0}{\s}>&=&\frac{4t_{i0}}{U}\p{i0}+O(t^3/U^3)
\ea
\ba
\sum_{\s}<\cd{i}{\s}\c{0}{\s}n_{0,-\s}>&=&\frac{2t_{i0}}{U}\p{i0} \nonumber \\
+\sum_{j,R_j}\frac{t_{iR_{j}}t_{R_{j}0}}{U^2}\{3\p{iR_{j}}-\p{R_{j}0}&-&\p{i0}+3{\it i}<\vec{S}_{i}(\vec{S}_{R_{j}}\times\vec{S}_{0})>\}
\ea

On a square lattice and in a state that doesn't break time reflection invariance\cite{T} ($<\vec{S}_{i}\cdot(\vec{S}_{R_{j}}\times\vec{S}_{0})>=0$) the second sum is (for different i):

\

\begin{tabular}{|c|c|}\hline
1&$4tt'\p{20}+2tt''\p{30}$		\\ \hline
2&$2(2\p{1}-\p{2})t^2+4\p{3}t't''	$\\ \hline
3&$(2\p{1}-\p{3})t^2+2(2\p{2}-\p{3})t'^2	$\\ \hline
4&2$(\p{1}+\p{2}-\p{4})tt'+2(\p{1}+\p{3}-\p{4})tt''$	\\ \hline
5&$(2\p{2}-\p{5})t'^2+2(2\p{3}-\p{5})t''^2	$\\ \hline
6&2$(\p{1}+\p{3}-\p{6})tt''		$\\ \hline
7&2$(\p{2}+\p{3}-\p{7})t't''		$\\ \hline
10&$(2\p{3}-\p{10})t''^2			$\\ \hline
\end{tabular}

\

Substituting these terms into A1,A2,A3 yields results of Section III.


\begin{references}
\bibitem{T} When this is no longer the case $<\vec{S}_{i}\cdot(\vec{S}_{R_{j}}\times\vec{S}_{0})>$ gives rise to $sin k_x$ ($sin k_y$) contributions to spectral moments resulting in $A_n(\vec{k})\not = A_n(-\vec{k})$.
\end{references}
\end{document}